# Laser Scanning Confocal Microscopy of Comet Material in Aerogel

MICHAEL GREENBERG[1*], DENTON S. EBEL[1]

[1]Department of Earth and Planetary Sciences, American Museum of Natural History
Central Park W. at 79th St., New York, NY 10024

--------------------------------------------------------------------------------------------
**Abstract --** We report non-destructive 3-dimensional imaging and analysis techniques for material returned by the Stardust cometary collector. Our technique utilizes 3-dimensional laser scanning confocal microscopy (3D LSCM) to image whole Stardust tracks, in situ, with attainable resolutions <90 nm/pixel edge. LSCM images illustrate track morphology and fragmentation history; image segmentation techniques provide quantifiable volumetric and dynamic measurements. We present a method for multipart image acquisition and registration in 3-D. Additionally, we present a 3D deconvolution method for aerogel, using a theoretically calculated point spread function for first-order corrections of optical aberrations induced by light diffraction and refractive index mismatches. LSCM is a benchtop technique and is an excellent alternative to synchrotron x-ray computed microtomography for optically transparent media. Our technique, developed over the past 2 years, is a non-invasive, rapid technique for fine-scale imaging of high value returned samples from the Stardust mission, as well as various other samples from the geosciences.
--------------------------------------------------------------------------------------------

## INTRODUCTION

Cometary material returned by the Stardust mission was captured into aerogel, an optically transparent porous low density $SiO_2$ solid, at a relative velocity of 6.1 km/sec. It is difficult to unravel the impact history of solid cometary material captured in low density material at hypervelocity. The 3D geometry of each Stardust "track" or hypervelocity impact structure, reveals the history of one impact event, acting as a timeline of particulate fragmentation and entry dynamics. We have utilized laser scanning confocal microscopy (LSCM) as a tool for imaging and analysis of these tracks in 3-dimensions, and at the highest possible spatial and contrast resolution. Application of benchtop confocal microscopy has returned fine scale observations of track morphology and deposition, unparalleled imagery of track structure in 3D, discovery of several new major fragments, and terminal particles, as well as discovery of entirely new tracks in aerogel keystones. We have also developed a system for correction of LSCM images using a theoretical point spread function and corresponding 3D deconvolution algorithm, prior to quantification of data. Here, we present the current state of techniques developed over three years. The techniques build on existing confocal principles developed primarily for the biological imaging community. Techniques presented here should apply directly to other optically transparent samples, specifically certain types of samples used in the geosciences.

Our investigation into 3D morphology and the formation history of Stardust tracks began with synchrotron x-ray computed microtomography (XR-CMT) (Ebel and Rivers, 2007). XR-CMT techniques were abandoned due to slow turnaround times, and relatively low resolution and visual contrast, and concerns over beam interaction with potential organic molecules in the Stardust samples. LSCM, on the other hand, is a rapid, benchtop alternative, performed at the American Museum of Natural History (Ebel et al., 2009; Greenberg and Ebel, 2009). 3D microimaging of whole keystones is actively pursued by several groups worldwide utilizing a variety of techniques. Kearsley et al. (2007) used confocal imaging to determine the structure of tracks, but the majority of recent published work uses XR-CMT (Nakamura et al., 2008; Tsuchiyama et al., 2009), but cannot provide the same level of detail and contrast as LSCM on entire tracks. SEM X-ray ultramicroscopy remains a possibility as well, and was discussed by Kearsley et al. (2007). LSCM is capable of images <90nm/pixel edge in the X andY directions and <350nm/pixel in the Z direction, while also remaining a non-destructive technique. We have developed a robust technique for imaging of whole tracks in full resolution, including procedures for tiling 3D image "blocks". We have imaged eight cometary tracks returned from Stardust, here concentrating on Track 152 (curatorial assignment: C2035,2,152,0,0) which represents the best of LSCM imaging capabilities. The scientific community as a whole will benefit from the availability of nondestructive 3D imagery of tracks prior to subsequent destructive analysis of individual particles.

Confocal imaging is widely used in the biological sciences, yet its adoption by the geological sciences community has been slow. An excellent reference on confocal imagery is the Handbook of Biological Confocal Microscopy, 3rd edition (Pawley, 2006b). While others have promoted the tremendous potential for confocal microscopy in the geological sciences (Makuo et al., 2009), the application of these techniques has mostly been limited to 1) 3D porosity measurements, and 2) volumetric measurements of oil in liquid inclusions (Menendez et al., 2001; Pironon et al., 1997). Practical limits on laser penetration depth constrain the abilities of confocal imaging on larger samples, which may in turn drive interest in alternative, widely available microtomography techniques. A typical confocal imaging system is limited to ~650 μm depth range, while an upgraded system with a pulsed, two-photon laser imaging setup could double this range, and would be required for imaging of samples thicker than ~1mm.

Previous LSCM studies in the geosciences have mostly investigated the pore structures and sizes of various materials (Petford et. al., 1999, 2001). These techniques involve the injection of optically transparent, fluorescent doped epoxies to fill the pores of materials. Material is then thinly sectioned and imaged in 3D (Fredrich, 1999). Due to recent advances in benchtop and synchrotron XR-CMT, and complexity of sample preparation, confocal analysis of porosity has been largely abandoned. Confocal analysis remains an excellent prospect for analysis of oil or other transparent media trapped in fluid inclusions (Pironon et al., 1997) as well as surface characterization in 3D. Accurate volumes and volume percentages of many structures can only be properly characterized in 3D. Lack of contrast in XR-CMT imagery, and oversubscription at synchrotron beamlines capable of imaging microscale structures leaves confocal as an attractive option for imaging many types of samples.

## METHODS

The principle of confocal imaging (Minsky, 1961) involves the use of a pinhole to block light from surrounding areas, focusing on a single point of focus rather than an entire field of view, as is the case with widefield microscopy (Fig. 1). This technique effectively removes light from out-of-focus points and planes, providing a crisp image with excellent resolution. The use of multiple lenses separated by a varying distance extends this principle to the third dimension. Laser scanning over the X and Y directions allows for imaging of 2D optical planes, and changing Z focus changes effective imaging depth within the sample (Inoué, 2006). Capturing 2D confocal planes at regular depth intervals gives a true 3-dimensionally imaged result. Note that the production of confocal images on a computer is the result of several interactions of laser light, first with reflectance or fluorescence of the sample to be imaged. The intensity of the returned image is recovered by photomultipliers within the microscope detector.

Images may be generated using a variety of laser wavelengths in either reflectance or fluorescence modes, in some cases using several laser wavelengths at once to capture a multi-channel image of several fluorescent effects. In contrast to classical optical microscopy, the only way to visualize confocal data is with computer software. Due to this and the large datasets created by this technique, the usability of confocal microscopy has been limited until the development of modern computers. Even today, the visualization and analysis of confocal imagery of the highest resolution can only be completed by high end computing devices.

### Equipment

All LSCM images were acquired at the Microscopy and Imaging Facility at the American Museum of Natural History, using a Zeiss LSM510 inverted confocal microscope, mounted on a Zeiss Axiovert 100 mount (Fig. 1). The entire confocal system is mounted on a passive air pressure stabilization table to minimize the effects of vibrations (e.g. subway) in the imaging process. Images are saved in a proprietary Zeiss .lsm file format, 8 or 12-bit depth. The .lsm format is a container for concatenated tiff stacks with extra metadata fields to hold confocal-specific data. The Zeiss system is attached to a 32-bit, dual core, Windows XP system. The 32-bit system cannot address physical memory sufficient to handle visualization and processing of the imaging data, thus all data is stream captured to hard disks before being moved to the storage area network at the AMNH. Post processing is performed on a 64-bit Windows XP computer with 24GB of ram and 4 processing cores. The Zeiss system runs Zeiss LSM version 3.2, a version of the software not originally intended for use with the LSM 510 system, thus some of the features present in software are not usable with this microscope system. Basic visualization of the image data is preformed with NIH ImageJ software (v. 1.43i) and associated ImageJ plugins written by various members of the community. Volume rendering is preformed with Bitplane Imaris 6.3. Subsequent deconvolution of LSCM data is done with SVI Huygens Professional 3.4.0. Our procedure for imaging, post processing, data re-assembly, and analysis uses these tools as described below.

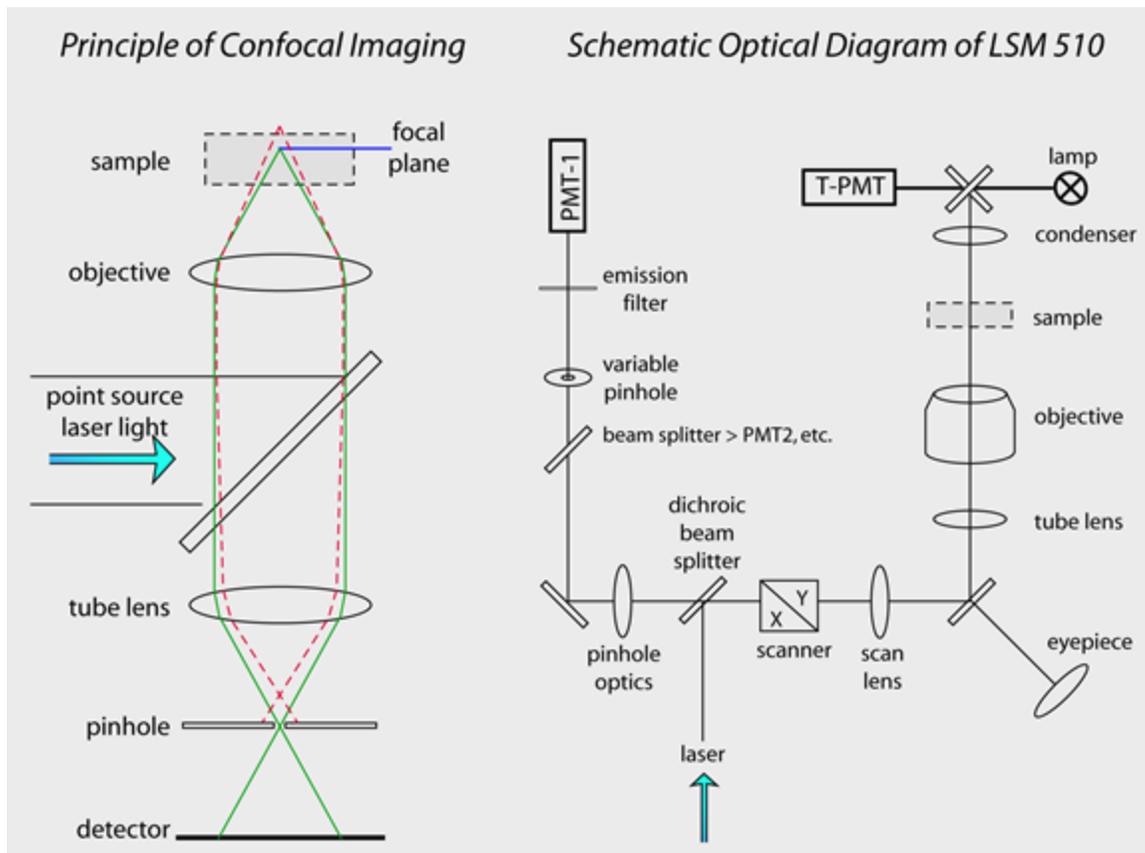

**Figure 1** - Left: illustration of principle of confocal imaging. Right: light path diagram for the Zeiss LSM 510 (PMT = photomultiplier tube).

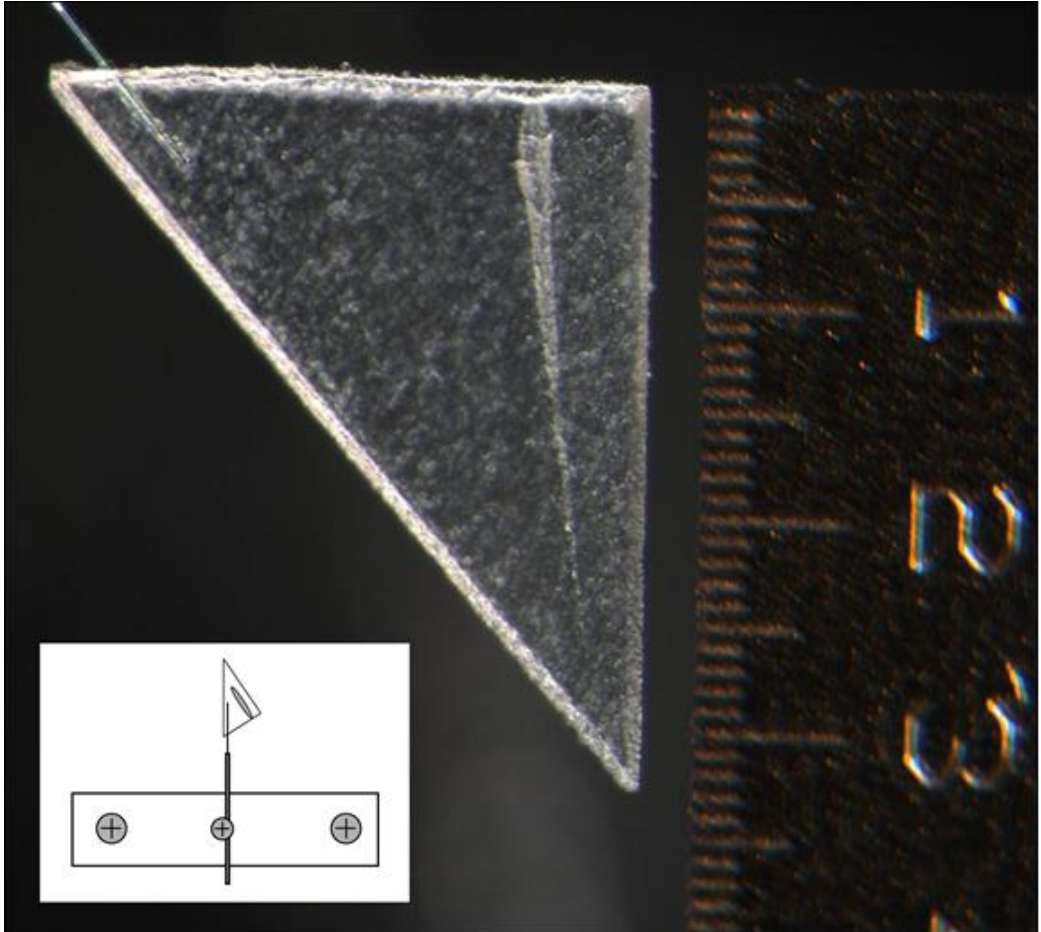

**Figure 2** - Digital photograph of Keystone 152 on its fork. Scale in mm. Image courtesy D. Frank, JSC. Inset: Schematic of new acrylic sample mount.

## Samples

Our use of confocal microscopy stems from a mandate to the preliminary examination team of the NASA Stardust mission. Comet Wild 2 dust particles and their tracks represent some of the most high-value samples in planetary sciences, being the only solid samples returned to earth since the completion of the Apollo program. The key benefits of confocal imagery in this case are its non-destructive nature, and its ability to image nanoscale structures in 3 dimensions. Cometary dust particles returned from the NASA Stardust mission were captured at hypervelocity speeds (6.1 km/sec) in 2 x 4 x 3cm deep aerogel tiles, creating cavities, or dust 'tracks' (Burchell et al., 2006; Tsou et al., 2003). Tracks are systematically created by the particles they contain and morphology of each track is unique to the original composition and fragmentation history of the impactor and resulting terminal particle(s) (Burchell et al., 2008). Whole tracks are extracted in triangular prisms of aerogel, called keystones (Westphal et al., 2004). Thus far we have imaged a total of eight stardust tracks, totaling five keystones. Keystone 152 is ~3.4mm in length and ~650 μm in thickness (Fig. 2), and contains only one track. The majority of this work concentrates on track 152, since it reflects our most current imaging techniques. Cometary Stardust track 152 (C2035,2,152,0,0) is a large carrot shaped "type

A" track ~2.618mm in length. It has a small bulbous portion ~300 μm long by ~200 μm wide and is the largest track we have imaged at the highest possible resolution.

Keystones are extracted and mounted using a "forklift," a fragile 25 mm glass rod with tines on one end that hold the aerogel keystone (Westphal et al., 2002). For additional safety, we have manufactured a custom mounting apparatus for use with forklifts. Our apparatus is similar in dimensions to a standard microscope slide but is custom milled from clean acrylic plastic. The keystone is mounted orthogonally to the long edge of the slide and is secured by two acrylic screws (Fig. 2, inset). Our custom mount is also designed for safe transport of the keystone. This apparatus can be used to perform several types of whole keystone analysis, including LSCM, synchrotron X-ray fluorescence (SXRF), and synchrotron X-ray diffraction (SXRD), all without handling the forklift rod itself.

| Objective | NA | Pixel Size | Field of View Width |
|---|---|---|---|
| EC Plan-Neofluar 2.5x | 0.075 | 0.610μm | 1249.3μm |
| EC Plan-Neofluar 5x | 0.16 | 0.381μm | 780.3μm |
| Fluar 10x | 0.5 | 0.122μm | 249.9μm |
| Fluar 20x | 0.75 | 0.083μm | 165.9μm |

**Table 1:** Objectives and numerical apertures (NA). Pixel size (X, Y edge length) and field of view width are calculated using optimal Nyquist sampling rates. Tiling of scans is necessary to image large structures at the highest possible resolution.

**Technique**

Due to the unique nature of aerogel keystones, there are many factors to consider when imaging samples from Stardust. First, small wafers of aerogel (<1mm thick) are incredibly fragile, and subject to brittle fracture by even very small forces. Extreme care must be taken when handling keystones and during imaging it is most important that the microscope objective not come in contact with the keystone. In the LSM 510 inverted microscope the optical objective is controlled by a fine scale motor and moves relative to a fixed stage. A first step is to set a maximum Z value for the objective to avoid possible collisions. A camera-based alignment solution is optimal. Due to the hydrophillic nature of aerogel, the use of oil immersion lenses is not feasible unless the sample is completely sealed in a housing. Aerogel readily acquires a static charge due to its high surface area, and a Polonium source for static neutralization is advisable.

Once microscope-specific calibration is completed, image orientation and acquisition can begin. The Zeiss LSM 510 is equipped with four lasers of different wavelengths: 458nm Ar, 488nm Ar, 543nm HeNe, and 633nm HeNe. The majority of confocal imaging in the biological sciences is preformed using fluorescent dyes to differentiate structures. The use of such dyes is strictly prohibited by the nondestructive nature of our procedure, and fluorescence imaging is further precluded by a lack of naturally fluorescent material in comet Wild 2 samples. There are indications that flight grade stardust aerogel fluoresces under ultraviolet light (Sandford, 2006), but the LSM 510 we use is not equipped with a UV laser source.

Confocal images are reflectance intensity images acquired using the 488nm Ar laser. This laser wavelength was chosen due to its superior reflectance intensity and for maximum resolution. A variety of lenses are used for imaging. Lower magnification lenses are used for whole keystone "field scans" and higher magnification lenses are used for tiled high-resolution scans of track details. A summary of dry microscopy lenses and parameters is presented in Table 1. Laser power is kept at a constant 75% to ensure 4.2A of current to the laser. These are optimal conditions for the LSM 510 setup as indicated by discussions with Zeiss representatives. The percentage of laser light used for imaging can be controlled by the LSM software and is kept at 11% to avoid oversaturating the detector, while preserving a maximum amount of reflectance data. Clipped images, which have contrast values beyond acquirable range, reduce the ability to perform post-processing. Scans using 8-bit acquisition mode record images in 256 grayscale values; 12-bit mode gives a superior 4096 grayscale values of digital contrast and is used for all of our recent imagery. Effective image contrast can be controlled through software using detector and amplifier gain settings. Detector gain should be as low as possible so the detector is not oversaturated (Pawley, 2006c). Our microscope was purchased in 1997. Newer confocal microscopes have greater light-path clarity and finer scale control over collection parameters, which produce clearer images. Newer equipment can also collect images in 16bit format, providing 16x the grayscale contrast for use in image analysis.

Optimizing X, Y and Z sampling rates is also a practical concern in confocal microscopy. Since LSCM images are created by scanning a laser over a surface one line at a time, it is important that samples are taken at an optimal interval to maximize the resolving power of the objective used. It is in the sampling rate that compromises may be made to reduce the size of total data collection, but post-acquisition 3D deconvolution requires raw images with proper sampling rates. A correct deconvolution cannot be run if the images are not properly sampled. The Nyquist theorem stipulates the sampling rate necessary to extract the maximum amount of image data (Pawley, 2006a).  To be safe, proper sampling in the X and Y directions should be performed at 3x the Rayleigh criterion or more; sampling at this rate is beyond the stipulations of the Nyquist theorem. Oversampling of the data can always be corrected later, but data missed by undersampling can never be corrected.

Correct sampling in the Z direction is determined by several parameters and is highly dependent on the size of the pinhole used. The confocal pinhole size is defined in airy units, $d_{airy} = (1.22 * \lambda * \text{magnification} * 3.6)/(NA)$, where NA is the numerical aperature and $\lambda$ is the wavelength. Use of a pinhole of 1 airy unit in size is optimal for high signal/noise ratio. Smaller pinholes produce thinner optical slices, but have less illumination of the sample. The correct axial sampling rate (on Z) is typically much larger than in the X and Y directions (Pawley, 2006a). For our highest resolution images the correct sampling rate in Z is 360nm, compared to 83nm in the X and Y directions. Since sampling rates may be different in the X, Y and Z directions, volume elements, or voxels, may not be cubic, but rather elongated on the Z axis. To achieve such high sampling rates the LSM 510 uses a digital zoom feature. A pair of focusing mirrors raster the scanning beam over a smaller area; typical digital zoom systems enhance the resolution of images by performing digital interpolation of imaged data. The size (bytes) of a correctly sampled data set may seem unwieldy, but it is crucial to ensure that deconvolution is completed correctly.

Although confocal microscopes are benchtop lab equipment, scan time constraints remain an issue. The LSM software allows the user to vary the per pixel dwell time.

Faster scans are better for alignment visualization, while much slower dwell times should be used for data acquisition. In general, a per pixel dwell time ~10μs is the point of diminishing returns. The software also allows the user to average the scanning data N times at point, line or frame intervals. We find that the averaging of every line twice, or "line-mean-2" scanning provides the best combination of speed and image quality. Additionally, contrast levels change with scan speed. After image position alignment, frames from the center of the image stack should be acquired at the desired scan speed, to determine the correct gain and offset values to avoid oversaturation or underillumination. Physical sample drift is also a concern when selecting scan speed. If a drift of more than a few pixels is to be expected, more lines or frames may need to be averaged to compensate for the drift effects. Small, uniform shifts in an entire frame are easily corrected in post processing, but random, small shifts throughout a hastily acquired image are nearly impossible to correct. A properly sampled scan with appropriate per pixel dwell times can take between 3-5 hours. Scanning of an entire track ~1mm in length can optimally be accomplished in less than 2 days. Imaging a structure of this size at the highest possible resolution requires the application of a scan tiling method that we have developed. Unfortunately the LSM 510 is not equipped with a motorized stage and queuing capability in the software, therefore sequential scans must be set up manually, at the microscope. The LSM 710 model, equipped with a motorized stage does not suffer from this defect. Our sample geometry and the working distance of the microscope objective limit imaging depth. The 20x lens (Table 1), which is most commonly used for high-resolution imagery, has a working distance ~600 μm, which is near the maximum penetration depth of confocal microscopy (Pawley, 2006a). In the case where structures are >600μm thick we have developed a procedure where the sample can be rotated 180 degrees and a complementary image is taken from the obverse side. Resulting 3D "blocks" are seamlessly stitched together in post-processing.

**3D Deconvolution**

3D deconvolution is a necessary procedure for the correction of LSCM data. Due to the diffractive properties of light around the confocal pinhole (Fig. 1), blurriness and elongation along the optical axis are almost always present in confocal imagery. This blurriness is systematic to every imaging setup and is highly dependent on many microscope parameters: laser wavelength, pinhole radius, sampling rate, slice depth, lens magnification and numerical aperture, as well as optical properties of the sample itself. The characteristic shape of the optical aberration is called the point spread function (PSF) of the optical system. Further complications in the PSF arise when the index of refraction of the microscope objective does not match the refractive index of the medium imaged. This refractive index mismatch causes an asymmetry in the PSF. Large mismatches in refractive indices create wildly elongated PSFs which are more distorted with deeper penetration into the sample. Deconvolution of these images is done separately with multiple PSFs applied at varying depths, sometimes upwards of 40 PSF models are used. Proper deconvolution of images increases the sharpness in the X, Y and Z directions, while preserving the intensity information of the original image.

There are three common methods for 3D deconvolution, the first being blind deconvolution, wherein no PSF is used, rather a guess is made by the software based on optical parameters and an iterative process is used to refine results (Holmes et al., 2006).

This solution is best applied on simple optical systems. The second method, which we intend to use in the future, requires submicron spheres of regular radius to simulate singular points. Imaging of these spheres in 3D constitutes a measured PSF (Juškatis, 2006). We have not yet been able to acquire suitable samples of submicron spheres embedded in either flight grade aerogel or melted silica glass. The third method, which we use extensively, involves the calculation of one or more theoretical PSFs for use in an iterative process (Cannell et al., 2006). Optical parameters must be determined precisely in this case because the PSF varies significantly with depth in the sample.

The Huygens Professional software makes deconvolution a manageable task, even on huge datasets. The choice of optical parameters for deconvolution is an iterative process, constantly in revision. Deconvolving 3D images of keystones is not a rudimentary task, for the optical properties of keystones are not constant nor are they easy to constrain. The majority of material in keystones is pristine aerogel, unaltered by the capture of comet dust. Flight grade aerogel has an average density which corresponds to an average refractive index of ~1.028, essentially that of air. The capture of cometary material produces heat and pressure on the aerogel, causing melting and compression, changing its density and thus its index of refraction. The optical properties of molten aerogel should be similar to an amorphous silica glass, which has a well defined index of refraction of 1.458. The density of compressed aerogel is less well defined, thus the index of refraction of these areas cannot be well constrained. Thicknesses of these compressed and molten aerogel components vary throughout the sample but should be no more than a few microns. Finally, the tracks contain cometary material, which is not optically transparent and appears highly reflective in LSCM images. Large cometary particles should be sheathed by molten aerogel and should behave as silica glass. This complicated optical system makes for a difficult deconvolution procedure. The deconvolution method we use assumes a refractive index of 1.028 in a classic maximum likelihood estimation algorithm (CMLE). This is an iterative method of deconvolution that may take many iterations before the stop criterion is met. We set a maximum of 150 iterations or a quality change factor of 0.01% as a stop criterion. Since the refractive index mismatch is set very low, a fixed PSF is used and deconvolution is processed on roughly cubic 3D sections or "blocks" of the image. The number of blocks is determined by the available RAM on the computational system. Systems with more RAM can deconvolve larger blocks and provide more accurate results, yet larger blocks take longer to deconvolve. One deconvolution run on a 2048x2048x300 pixel image stack could take between 4 and 8 hours depending on the complexity of the PSF. It is important to note that deconvolution should always be run on raw data; images that have been processed even slightly will not deconvolve properly. The developers of the Huygens software, requiring frequent communication, are consistently optimizing the deconvolution algorithms. Due to the complexity of our optical system we have been working on several radical deconvolution procedures, including a "mixed deconvolution," where a few iterations of alternating PSFs are used to simulate a hybrid PSF. We hope to report on the results from this procedure soon. The results of properly sampled and properly deconvolved imagery are far more detailed than optical imagery, and make 3D reconstruction and segmentation of objects and surfaces simple, due to reduced uncertainty in the data. A comparison between images is presented in Figure 3.

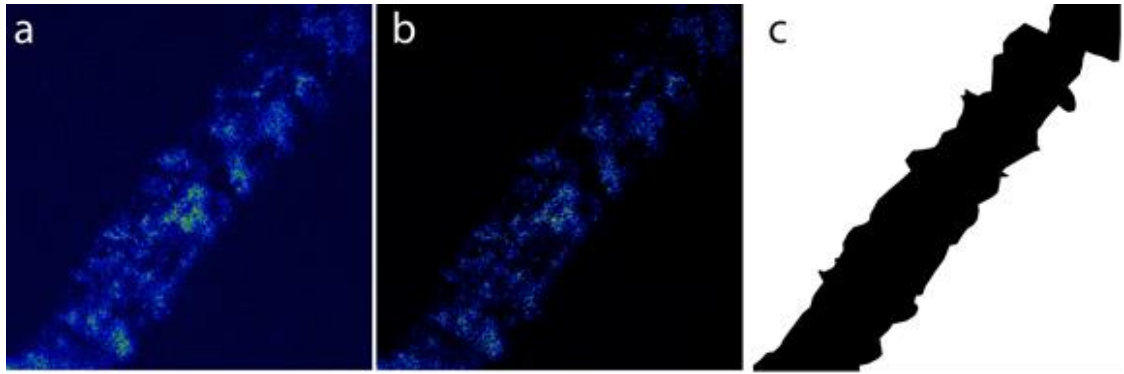

**Figure 3 -** Comparison between (a) raw image, (b) deconvolved image, and (c) segmented image.

## Post processing & Analysis

Due to the complexity of our images, typical automated segmentation techniques cannot be applied. Even in properly deconvolved images there is no one grayscale value which defines the boundary between track interior and clean background aerogel. Threshold values may vary in sequential slices because laser attenuation varies with depth. For these reasons all data is manually segmented. Manually segmenting ~60GB of data for one track is a highly tedious task and we are actively working on a means of automated segmentation. Segmented data is binary, with the track interior having a grayscale value of 0 and the exterior having the maximum possible value. Segmentation routines are carried out with procedures for NIH ImageJ software.

Quantitative analysis of huge datasets can be tricky and computationally intensive. Track images that are too large to hold entirely in memory must be analyzed one image stack at a time. This requires precise knowledge of the relative positions of the two stacks which can be tracked using a small dot or glyph to indicate one equivalent point in two overlapping images. Absolute transformations can be used any time after, if relative positions of each overlapping image pair are known. To this end, we have developed a rapid image quantification toolkit. Once images are segmented and thresholded to binary, they are rotated to a vertical alignment and IDL routines are used to calculate cross sectional area, cumulative volume and a parameter we define as skeweness, all as a function of increasing penetration depth along the track. Skewness is defined as the deviation of the centroid of a single cross-sectional slice normal to a straight line drawn from the center of the entry hole to the center of the terminal particle. This parameter can be used to quantify nonlinear motion of the impactor as a function of time and position.

| Scan | Lateral Resolution | Vertical Resolution | Slices |
|---|---|---|---|
| Field | 5.0 μm | 8.0 μm | 123 |
| 1 – 15b | 0.082 μm | 0.360 μm | various |
| 16a – 17b | 0.130 μm | 0360 μm | 400 |

**Table 2:** Scan resolutions. Slices for scans 1 – 15b vary with track thickness. Scan 1 is of the terminal particles and 17a and b are of the entry area.

## RESULTS AND DISCUSSION

The techniques described above were used to scan and process LSCM data from track 152, a ~2680µm structure. Physical presence at the microscope is required to load sequential scans, but 24-hour access is not feasible, so data was collected discontinuously over 9 days. A total of 23 3D scans were completed, one field scan of the entire keystone, and a 22-part mosaic of the track at the highest possible resolution. Eight of these images are members of scan pairs, stacked in the Z direction, and are labeled a and b respectively. These datasets were broken up due to the file size limitations imposed by 32-bit operating systems and the given file size limitation of the .lsm container. A summary of optical parameters used to image track 152 is in table 2. After processing all data was converted to 16-bit tif sequences, and imaged "blocks" stacked in the Z direction were recombined as one large tif sequence. Intensity values in 12-bit images are multiplied by 16, stretching the contrast to fit 16-bit depth.

Qualitative image results of track 152 indicate the presence of two large terminal particles, within 100µm of each other. A third terminal particle of notable size is apparent about 2/3 down the track. Many track features and varying morphologies can be observed in the images. 3D data cannot be fully appreciated on a 2D media, but to help visualize the detail of 3D data a mosaic of successive Z slices of the entry area of T152 is presented in Figure 4. We see characteristic fragmentation of the aerogel near the entry hole, with large "cones" of torn aerogel material extending radially and orthogonal to the track direction (Fig. 4). Also observed in the images is a periodic deposition profile in the stylus portion of the track, similar to the "rifling" observed by others (Horz et al., 2009; Nakamura-Messenger et al., 2007). A full resolution track map of T152 is over 30,000 pixels in length, so a highly deresolved image is presented (Fig. 5) for reference. Image data from track 152 has not been fully segmented by hand, so bulk statistics cannot be presented here. Presented instead is an image of previously imaged track T128a and its corresponding cross-sectional area data (Fig. 6, Greenberg and Ebel, 2009).

## CONCLUSIONS

We have outlined an experimental procedure for laser scanning confocal imaging and analysis of whole cometary tracks from the Stardust comet sample return mission. We have discussed possible further use of 3D confocal imaging in the geosciences and encourage the prospect of such applications. Our 3D confocal imaging technique while not directly applicable to all microimaging cases should provide a set of guidelines for proper imaging. The importance of proper spatial and contrast sampling cannot be stressed enough, for proper imaging technique is a precursor to proper analysis. Total non-destructive imagery of cometary tracks from Stardust benefits all members of the scientific community. Such imaging will be required for microphysical modeling of the impact histories of comet samples, and inferences about their original volatile content. The 3D imagery can be stored and recalled long after the destruction of the sample or the completion of other analyses. The combination of 3D spatial and morphological data from LSCM with other non-destructive analyses such as SXRF and SXRD (Flynn et al., 2006; Lanzirotti et al., 2008) enables a near-total track characterization.

The microimaging technique described here can be used for nanoscale imagery, with pixel resolutions <90nm achievable in the X and Y directions. Corrections of raw data by 3D deconvolution using theoretical point spread functions can be used to achieve greater accuracy in parameters measured from confocal images. The techniques used for imaging and data processing will continue to be refined, but after two years of experience, these techniques are at a very mature state.

*Acknowledgements*

We would like to thank Mike Zolensky for encouraging this work and recognizing the importance of full 3D imaging. We also thank Rebecca Rudolph, Emily Griffiths and Matthew Frenkel for their assistance with the LSM 510 system, and summer students, Stacy Ramcharan (Columbia University), and Peter Hein (Riverdale Country School) for their data crunching and preparation efforts. MG and DSE were supported by the American Museum of Natural History, and National Aeronautics and Space Administration grants NNG06GZ42G and NNX09AC31G.

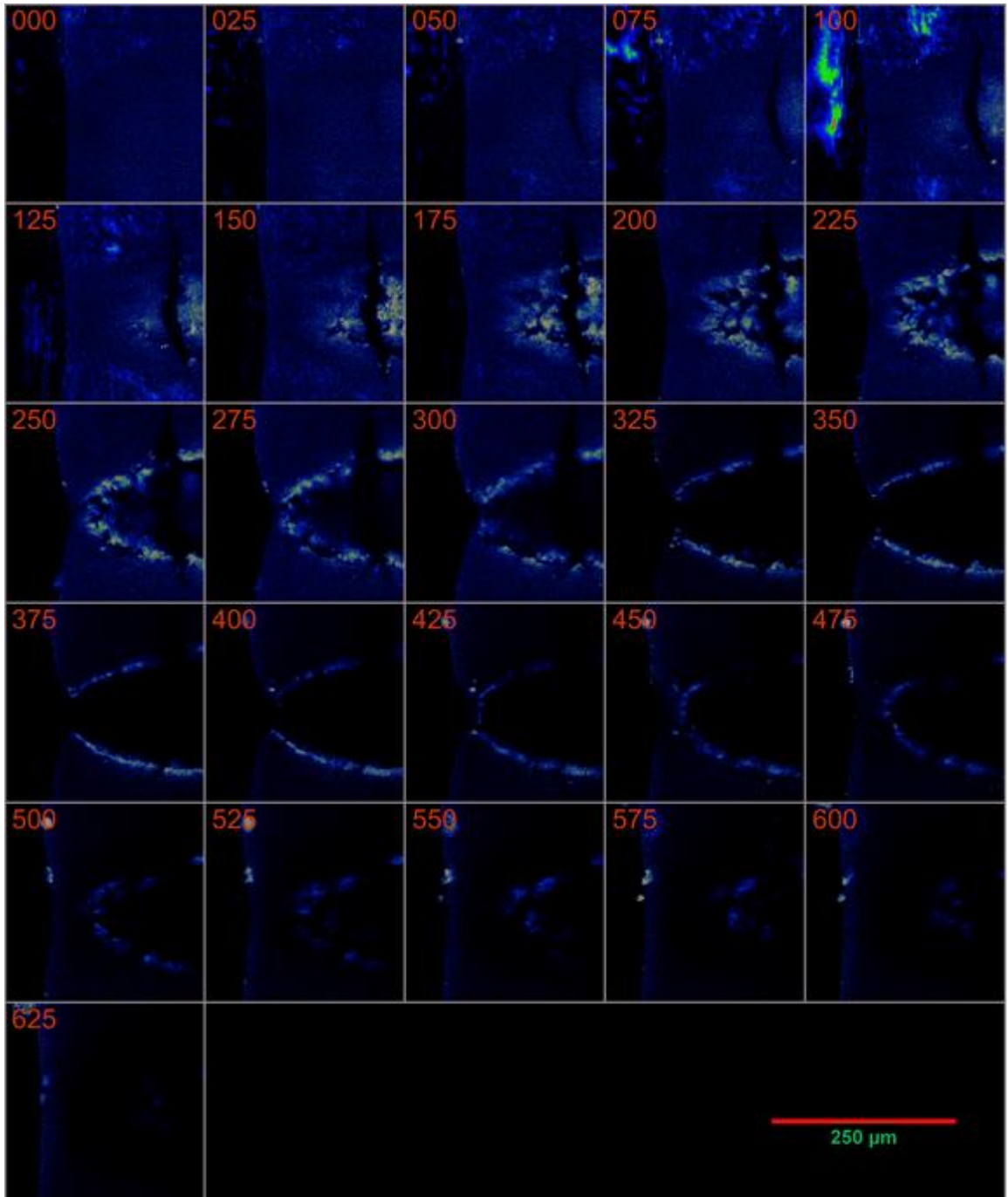

**Figure 4 -** Slices of bulb section of T152, 12-bit images. Every 25th slice is shown in this mosaic of the entry area of T152. A contrast stretch is applied where brighter areas indicate regions of higher reflectance.

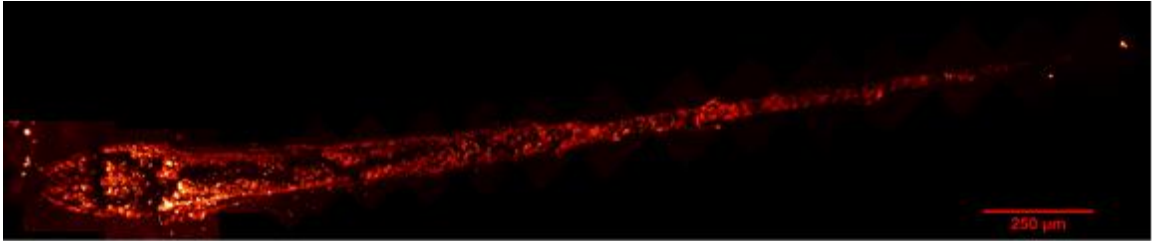

**Figure 5 -** Map of T152, 2D maximum intensity projection of deconvolved data.

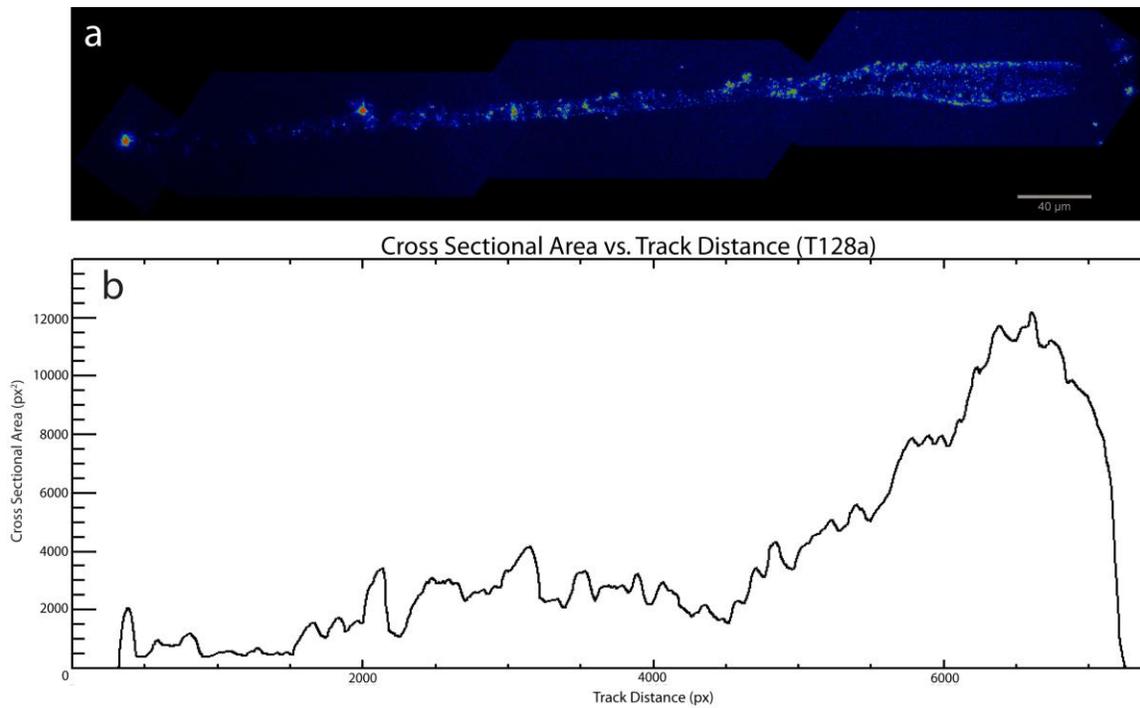

**Figure 6 -** Image of T128a (a) and corresponding quantified cross sectional area data (b). T128a is a ~600 μm long carrot shaped track. Image is 8-bit, deconvolved and stitched.